# Understanding Context to Capture when Reconstructing Meaningful Spaces for Remote Instruction and Connecting in XR

Hanuma Teja Maddali

Department of Computer Science, University of Maryland, United States, hmaddali@umd.edu

Amanda Lazar

College of Information Studies, University of Maryland, United States, lazar@umd.edu

Recent technological advances are enabling HCI researchers to explore interaction possibilities for remote XR collaboration using high-fidelity reconstructions of physical activity spaces. However, creating these reconstructions often lacks user involvement with an overt focus on capturing sensory context that does not necessarily augment an informal social experience. This work seeks to understand social context that can be important for reconstruction to enable XR applications for informal instructional scenarios. Our study involved the evaluation of an XR remote guidance prototype by 8 intergenerational groups of closely related gardeners using reconstructions of personally meaningful spaces in their gardens. Our findings contextualize physical objects and areas with various motivations related to gardening and detail perceptions of XR that might affect the use of reconstructions for remote interaction. We discuss implications for user involvement to create reconstructions that better translate real-world experience, encourage reflection, incorporate privacy considerations, and preserve shared experiences with XR as a medium for informal intergenerational activities.

## 1 INTRODUCTION

Extended Reality (XR) technology has seen increasing research and commercial interest as a medium to augment remote collaboration. When compared with traditional 2D media, XR provides a more effective way for users to perform spatial referencing and demonstrate actions remotely [45]. This can especially be helpful during collaboration and expert-guidance for physical tasks in professional settings (e.g. surgical training [20,57,95], field servicing [100]). The increasing access to XR devices is also leading to a growing body of HCI work on system building and interaction design for informal settings with remote friends and family. These works can involve, for example, games [2], learning [82], and other general-purpose telepresence applications [47,48]. Virtual representation of the users, their surrounding physical spaces, and objects in XR is an important element of this interaction design for collaboration or instruction. Recent advances in creating high-fidelity 3D environment [102,103] and avatar reconstructions [38,101] for XR devices are allowing remote collaborators to explore and obtain better spatial and semantic information from a remote environment.

The HCI community has explored various modes of representing real environments and objects as 3D models in XR to improve collaboration in remote guidance tasks [19,79,82]. These systems often focus on representing the activity space based on the constraints of the technology and designers' understanding of activity-specific interactions that need to be supported for guidance. For example, some objects might be explicitly prioritized for the demands of the activity by system designers during 3D reconstruction (e.g. high fidelity hand pose reconstruction during surgery [13]). In other works, often with room/table scale remote XR systems, the entire room/table volume and its objects are reconstructed by default [48]. There is also a focus on understanding how to create sensory affordance for the reconstructions comparable to their physical counterparts [70,71] (e.g. tactile sensations for virtual surfaces [75]). This can be useful for activities requiring precise sensory input for skilled physical tasks (e.g., cutting through material for surgery). This designer-led approach of creating virtual representations of activity spaces by capturing activity and sensory context might be optimal for professional settings with standardized practices. However, informal activities with friends or family can take place in settings where virtual representation of spaces and objects in XR needs to capture more personalized meaningful context. Affordances for virtual objects that enhance social connectedness between group members through "connecting" interactions (e.g. sharing virtual replicas) can be a priority to aid remote informal collaboration by building common ground [6,15]. But there is a missing understanding of how designers can translate the social context of real physical objects to their virtual representations and how this translation might affect user experience. Our work aims to better explore and understand this social context, in terms of expected social affordances, behaviors, and interactions, which reconstructed spaces and objects need to capture to be meaningful, for example, when connecting with a remote loved one. We take the case of hobby instruction in a closely related group (e.g., family, friends) of practitioners as an example of an informal setting for XR with the following research questions in mind:

- **RQ1**: What is important social and sensory context that can influence informal XR users' remote experiences with virtual representations of real activity spaces and objects?
    - **A.** What context is important to users for learning experiences with reconstructed spaces and objects?
    - **B.** What context is important for connecting interactions with reconstructed meaningful spaces and objects?
    - **C.** What context is important to users when sharing reconstructed meaningful spaces and objects?
- **RQ2**: What are some design considerations for XR to support the values and goals of an informal intergenerational group interacting with the virtual representations?

We choose gardening as an example hobby where social interaction for instruction or connection is often intergenerational, community or family oriented [39,93] thus providing rich context for physical spaces and objects associated with the activity. Although there are varying perspectives on technology inclusion in the space, socio-technological approaches that enable practitioners to learn from others have been suggested in the past as a fruitful research direction [26,39,41,46]. We conducted a study with eight groups of 18 intergenerational participants to evaluate an XR prototype for remote instruction scenarios in gardening. Our prototype allowed participants to view and interact with pre-created 3D models of meaningful real areas and objects chosen by them from their gardens. This allowed us to identify context that could be important for the models to augment instructional or connecting interactions over XR.

Our paper makes four contributions. First, we discuss if 3D models of a real activity space in XR can better support meaningful reflection during remote instruction (**RQ1A**). We note possibilities for reflection with different approaches to capturing context and constraints on using real-world-experience to understand the reconstructed spaces. Second, we provide an understanding of context that can capture social relations and emotional context (**RQ1B**). We discuss the value of virtual reconstructions to preserve shared memories and as mementoes compared to physical artifacts. Third, we discuss potential spatial privacy considerations for the process of creation and social sharing of virtual reconstructions (**RQ1C**). Finally, we discuss how our findings on creating and sharing virtual reconstructions of physical objects and spaces could apply to other intergenerational activities and provide meaningful directions to explore when designing informal or casual XR (**RQ2**).

## 2 RELATED WORK

### 2.1 Virtual representation of physical spaces and objects for remote guidance in XR (RQ1)

Past work on remote guidance in XR has covered virtual environment representations that include live or pre-captured 360-camera media [32,40,77], live or pre-rendered 3D reconstructions [20,48,79,82], and using virtual proxies for objects [17,45] in the environment. The design considerations that are highlighted in these works are often related to the relation between user experience and the context (e.g., visual features, physical manipulability) that is captured by different virtual representations.

*2.1.1 Context prioritized when virtually representing real environments*

Previous HCI research on reconstructing the user's physical location as a virtual environment has predominantly focused on an understanding of sensory context for reconstruction. This includes capturing and matching sensory context in terms of physical affordances (e.g. tactile sensations) of virtual objects with those of actual objects in the VR user's environment [70,71]. This can help the user with safe navigability while wearing a headset, and for experiencing interactive and sensory-realistic virtual objects and proxies remotely. For example, the bumpy surface of a rock can be approximated using a vibrotactile actuator [75]). This matching of affordances of the virtual reconstruction or proxy with their physical counterpart is an important design consideration for the psychological feeling of presence in a remote environment and maintaining sensory coherence with the physical environment around the user when exploring a mixed reality environment [73] (e.g. have similar tactile sensation [28]).

Researchers have also proposed abstracted virtual proxies by capturing a set of affordances (e.g. hand manipulability [45,58]) as context that is convenient for the user's intended tasks. Radu et. al [58] provide an example of virtual proxies for a virtual makerspaces that could capture context such as sensor data from their physical onsite counterparts, a model of their physical interactivity, and even discuss controlling these physical objects remotely through their proxies. There also has been substantial past research, and re-emerging interest in commercial XR, on creating 3D digital twins of environments or systems that can act as simulation proxies for evaluation, and even educational tools for their physical counterparts (e.g. rockets, classrooms, theatres) [1,47,64,68]. Design work using XR for cultural heritage preservation has focused on instructional affordances of virtually reconstructed objects (e.g. 3D scans of museum artifacts [56]) and digital twins of historic sites [9,62,63] and how they could engage the learner to consider their meaning. For researchers, and even a general audience, it can allow them to construct historical narratives by viewing and interacting with artifacts from all over the world in a single virtual "place" [84].

Commercial availability and capabilities of VR devices have increasingly improved for casual users. However, there is still a gap in current literature is a user-centered understanding of context beyond sensory features that is needed to create meaningful reconstructions for informal settings. A growing body of work has proposed understanding social context, in addition to sensory context, when interacting in XR with virtual representations of real environments in informal settings [40,85,90]. Maddali et. al suggest investigating how an XR environment might lead to additional affordances with making meaningful physical objects shareable and interactable virtually (e.g. arranging or reshaping together) and enable affective "connecting" interactions in hobby learning settings [40]. Based on their analysis, we argue that it is important to understand the breadth of social affordances for virtual objects in XR that enable connecting interactions, in addition to instructional (e.g., building observation skills), and physical affordances (e.g., being graspable). Our work provides a glimpse into what could be meaningful social context for 3D models of real spaces in informal learning scenarios. We also discuss approaches to capturing this social context for XR.

*2.1.2 Designer-led versus user-centeredness in identifying meaningful context for virtual representation*

Virtual representation of real objects that can be used as physical props in a collocated interaction is important for remote instruction and collaboration scenarios [48]. We find that past works that use virtual proxies of physical objects in XR to explore remote interaction workflows [45,48,51] often use a designer-led understanding of context to capture through the proxies. In room-scale telepresence it is assumed that all the objects in the room are reconstructed by default [48]. Some works include pre-created [17] or live [92] virtual proxies of objects important for instruction. The presentation of proxies can also be prioritized based on importance to instruction (e.g. higher hand resolution for remote surgery guidance [13]). However, these works do not discuss/utilize any structured user-centered approach to identifying the objects and the surrounding sensory and social context that need to be represented in XR (e.g., incorporating privacy preferences). Our work extends this space by identifying affordances for reconstructed spaces that arise when meaningful objects and spaces are intentionally selected with participant involvement. While we do find work that explores privacy of person [47,60], privacy of space and the concerns of the user when creating or sharing reconstructed objects or environments is rarely discussed. Wang et. al point out the importance of acknowledging such concerns when users create and share 3D reconstructed moments [85]. We add to this XR-privacy literature by looking at context dependent concerns and perceptions around creating and sharing virtual reconstructions of environments. We also present how the nature of the activity might influence how context is captured for the reconstructed models.

## 2.2 Intergenerational XR for meaningful informal settings (RQ2)

HCI researchers have often contextualized designing remote intergenerational systems within a trend of increasing social isolation due to several factors like the dispersed and nuclear distribution of families [61]. Pertaining to groups involving older adult users, the focus has been on use-cases that combine gaming in some form with storytelling [25,36,59]. Game-based approaches have some benefits, for example, as a way to combine light exercise with social interaction [33,49,91]. However, there might be wariness among older adults of prototypes and game-like activities that could be perceived as children's activities or unproductive [42]. We see some HCI researchers taking this into account by avoiding associations with gaming when prototyping for intergenerational experiences or for study procedures [8]. This consideration appears when the objective for using XR is an exchange of knowledge, for example, of traditional culture between older and younger users [14]. Positioning older adults as keepers of family history and younger people as memory triggers [36] for intergenerational storytelling and reminiscence applications [7,12,36] has seen a positive user reception in past studies. However, these studies used traditional 2D media and objects. Our findings provide an understanding of how experiences with 3D models of real places and objects in XR would be perceived relatively.

Established industry players like Microsoft [48] have shown visions of using 3D reconstructions of real spaces for casual XR outside workspaces. Meta has notably implied intergenerational use-cases for XR telepresence where you could watch sports *"together from a 3D model of his dad's apartment"* [31]. However compelling use cases, outside of gaming, for a more diverse demographic of casual users remains to be implemented and adopted. To expand to a use-case that might fit a different set of interests, we chose a popular activity linked to leisure and informal learning especially as a hobby activity in an intergenerational setting [40,41]. Gardening has a highly multigenerational demographic with older adults (65 years and above) being the largest age group [93] and most active volunteer group at extension master gardener programs [16]. In this work, we discuss intergenerational perceptions of social connection in XR around hobby learning as a meaningfully perceived use-case. We link these perceptions with experiences of interacting with 3D models of real spaces and the context that they should capture. Past work has also noted that prototypes and use-cases for XR can often be biased by a younger user and designer centric approach [65,66]. This can resulting in an experience asymmetry [44] for one of the age groups. In our study we note how this, and other factors could have affected perceptions of the state of XR and its future possibilities for intergenerational interaction.

## 3 STUDY DESCRIPTION

### 3.1 Participants

We recruited a total of 18 participants (avg. older adult age = 69.7, avg young adult age = 45.8, avg teen age = 14 years) (Table 1) and conducted eight sessions (Table 2) of our user study with six pairs and two groups of 3 participants. The pairs and groups included older adults (aged 65 and above), and family members, friends, or acquaintances who were atleast one generation apart (e.g., P6 was a friend of P5's daughter) with an avg age difference of 31.5 years. 13 participants described memories of learning hands-on in the garden with their family by spending "summers with my aunt(P1), my grandmother" (P2) or great-grandmother in P17's case. In four groups, (S1, S2, S4, S7), younger participants' childhood experience with gardening was in the presence of their older study partners. Beyond leisure, it was a necessary skill for P6 who "couldn't afford (veggies) to eat, so we had to plant (gardens) to eat". A few others (P5, P10, P12, P17) described teaching themselves later in life through online resources or professional programs while encouraging other young members of their family to help in the garden (P12). P7 (younger) and P10 (older) had been kindling their interest in gardening outdoors with the help of their study partners. This qualitative and quantitative profile of our participants allowed us to collect rich data on the intergenerational social experience with our prototype.

In all the pairs or groups, there was a significant, self-described difference between the members in the level of experience with gardening. Participants included four novices, 5 certified master gardeners, 5 experienced gardeners, and 3 "professionals" who were farmers (P3) and academics in agriculture and biology related fields (P6, P14). However, even here, experience between participants should be considered relatively. For example, P8 had substantial experience with farming compared to P7 and P15 was considered more knowledgeable than P14 who was a professional. The older participants were not necessarily the more experienced gardeners. For example, P1, P5, and P10 were either novices or less experienced than their study partners. This diversity in experience levels allowed us to collect feedback for our prototype from an instructional perspective.

In two of the sessions (S3,S4), participants were in different cities/states and were truly remote. For the other sessions, we simulated a remote setting by having the participants in separate locations (e.g., a participant in house and their partner in the garden). In four sessions, all remote/simulated-remote group members were familiar with the garden locations chosen for the study and reconstructions since they visited the location frequently. P6 mentioned having a general overview of the location for S3 but never having been inside it. In S4, P7 and P8 mentioned being introduced to P9's backyard only prior to the study over video chat. Only two sessions involved one group member viewing the selected garden location for the very first time (P13 in S6, and P15 in S7).

Table 1: Self-reported participant information

| Participant | Expertise | Age | Sex | Ethnicity | Relation between Participants | Prior AR/VR usage |
|---|---|---|---|---|---|---|
| P1 | Experienced | 65 | F | African American | P1 is P2's Aunt | Yes |
| P2 | Master Gardener [a] | 54 | F | African American | | None |
| P3 | Professional [b] | 74 | F | White | P3 is P4's Grandmother | None |
| P4 | Experienced | 15 | F | White | | Yes |
| P5 | Master Gardener | 70 | F | White | Good friends through P5's Daughter | None |
| P6 | Professional | 34 | F | White | | Yes |
| P7 | Experienced | 70 | F | African American | P7 is P9's Aunt. P8 is P7's Husband. | None |
| P8 | Experienced [a] | 74 | M | African American | | None |
| P9 | Novice | 41 | F | African American | | Yes |
| P10 | Novice | 67 | F | African American | P10 is P11's Mother | None |
| P11 | Master Gardener | 45 | M | African American | | None |
| P12 | Master Gardener | 67 | F | African American | P12 is P13's Mother-in-law. | None |
| P13 | Novice | 39 | F | African American | | Yes |
| P14 | Professional | 54 | M | Latino | P15 is P14 and P16's Neighbor. P14 is P16's Father. | Yes |
| P15 | Experienced | 69 | F | White | | None |
| P16 | Novice | 13 | F | Latina | | None |
| P17 | Novice | 54 | F | White | Good friends and former neighbors | None |
| P18 | Experienced | 71 | F | White | | None |

[a] Participant also had substantial farming experience. [b] Participant is also a Master Gardener

Table 2: Session information

| Session | Participants | Location(s) explored for garden walkthrough (Session Part 1) | Site Familiar to all participants | Session type | Models created for XR prototype (Session Part 2) |
|---|---|---|---|---|---|
| S1 | P1, P2 | P2's home garden | Yes | Simulated remote [a] | P2's backyard & decorative area in front yard |
| S2 | P3, P4 | Family farm | Yes | Simulated remote [a] | Unplanted farm plot & decorative area |
| S3 | P5, P6 | P5's home garden | No (P6) | Remote [b] | Fenced area with garden beds in P5's backyard |
| S4 | P7, P8, P9 | P9's backyard, P7 and P8's home garden | No (P7 and P8) | Remote [c] | Unplanted garden bed in P9's backyard |
| S5 | P10, P11 | P11's home garden | Yes | Simulated remote [a] | Two unplanted garden beds, & herb garden |
| S6 | P12, P13 | P12's community garden | No (P13) | Simulated remote [a] | Herb garden, butterfly garden, & memorial area |
| S7 | P14, P15, P16 | P14's community garden | No (P15) | Simulated remote [a] | Raised bed area, terraced slope, & bee house |
| S8 | P17, P18 | P18's home garden | Yes | Simulated remote [a] | P17's house garden bed, P18's back & front yards, & P17's comfrey plant at P18's house |

[a] Participants alternated between mobile phone (for onsite AR) and Oculus Quest headset (when simulating remote VR) . [b] P5 on Tablet (AR) and P6 on VR headset. [c] P7 and P8 on mobile, P9 on VR headset

### 3.2 Procedures

Each session was 90 minutes in duration and had two parts following the timeline in Figure 1. The onsite participants walked the researcher and the remote participants through a garden area preselected by them over video chat in part 1. This was followed by participants evaluating the XR remote instruction prototype in the expert or novice's garden in part 2. A short pre-study session was also conducted either remotely or via email instructions where videos and photos from the selected garden site were collected by the participants or by the researcher with participant guidance. These were used to create 3D models of areas and objects in the participants' garden to be used in part 2 of the study.

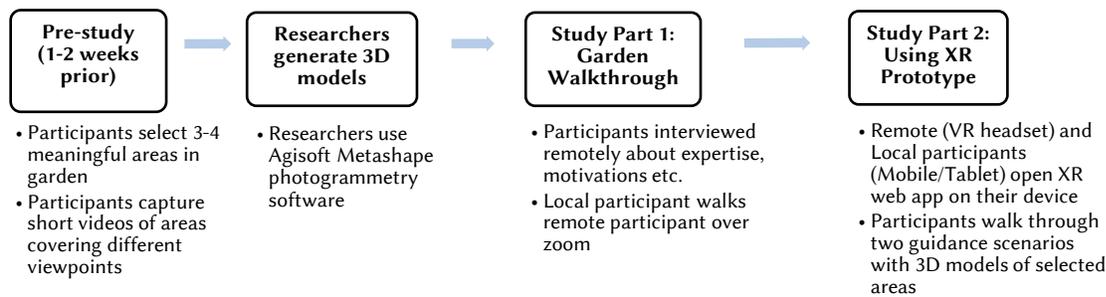

Figure 1: Summarized timeline for each participant session.

*3.2.1 Pre-study session*

Around 1-2 weeks prior to each session, the participants were instructed to choose 3-4 small areas (around 10ft by 10ft) in their garden that could be meaningful or interesting for them and/or their partners. These were areas that could be more familiar to them, have some objects/plants/features that they could use to teach something, or just might be enjoyable to be in. For each of these areas, the participants started capturing a 1–3-minute video by first standing just outside the boundary of the plot and doing a slow walk around its boundary while keep the camera pointed towards the center of the plot and coming back to the starting location. During the walk, the participants were suggested to 1) keep their camera movements smooth and slow to avoid motion blur, 2) try ensuring the camera's view is unobstructed by, for example a fence or tree and, 3) occasionally vary height and distance of the camera from the area to get a more comprehensive view of the area/object. An example video shot following these instructions was provided to the participants as a reference for the kind of video to be created. The researchers sampled the video frames and used 100-300 frames to generate each 3D reconstructed area/object using the Agisoft Metashape photogrammetry software [104].

*3.2.2 Part 1: Remote garden-walkthrough session*

In this part of the study, participants were first interviewed about their experience level, familiarity with the pre-selected garden area and objects, motivations for gardening individually and together (if done in the past). They then explored and conversed about the pre-selected garden areas and objects using conventional video chat as though they were remote from each other (in simulated remote sessions). The participants identified what components of those physical garden areas were meaningful to them individually and collectively (Figure 2.) and gave us a sense of how these components could be utilized by them for instruction or to feel connected. This session also helped establish a shared gardening experience that the participants could refer to compare their social experience with the XR prototype in Part 2. This approach was also intended to be useful for the participants to reflect on the idea of remote instruction during gardening session and provide them with a baseline on interaction using conventional video communication.

*3.2.3 Part 2: XR prototype evaluation*

On completing the gardening session, participants evaluated an XR remote instruction prototype informed by design considerations and a scenario for garden planning identified from past work [40]. The objective of this part of our sessions was to understand how the objects and areas from the practitioner's activity space identified through the gardening session can be represented and even augmented with meaningful context for instructional and social connecting interactions in XR. The prototype provided examples for virtual objects and environmental elements in XR that can be used by the practitioner for instructional or connecting interactions. The session will be followed by a semi-structured group interview with the expert-novice group to help evaluate the prototype.

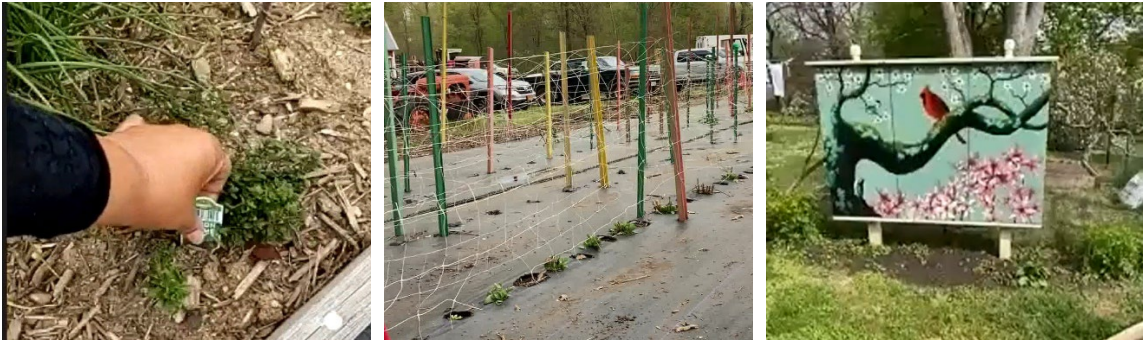

a) P12 showing P13 some herbs that she might like to cook with

b) P4 walking through a plot that she helps P3 with.

c) P5 showing P6 some artwork her neighbor made for their garden

Figure 2: Examples of meaningful objects and areas discussed during the garden walkthrough. Participants focused on objects that might be meaningful for their utility in activities they like, instructional value, or aesthetic reasons etc.

### 3.3 XR remote instruction prototype

Our XR prototype is a multi-user system through which participants could together interact with the photogrammetrically generated 3D models of pre-selected areas and objects from the selected garden spaces and gain familiarity with XR environments. It is an HTML and JavaScript-based webapp using the networked A-FRAME [105] package. Networked A-Frame uses WebRTC and WebSocket connections for audio-visuals and syncing remote users. Users can simply open a link to the prototype website on a browser in their mobile/tablet/VR headset, like a zoom meeting link. By default, A-FRAME creates buttons on the browser UI to toggle between AR, VR, or in-browser modes. This allowed participants to use devices that were immediately accessible without restricting them to headsets for AR/VR. It also addressed health concerns around sharing headsets with other groups in the current COVID pandemic. In our sessions, participants onsite on the garden used the prototype in AR mode on a mobile/tablet and remote (or simulated remote) participants used the VR mode with the Oculus headset.

We used two scenarios appearing in past work [40] as evaluation scenarios for the prototype that illustrate instructional and connecting interactions in XR for gardening. In these scenarios, the onsite participant uses the AR mode of our prototype on their device (a tablet or mobile phone) to view a 3D reconstruction of the terrain of the garden while the remote participant views the model using the Oculus Quest headset in VR mode. In simulated remote setting, participants played both the onsite and remote roles alternately.

#### 3.3.1 Scenario 1 – Gauging the slope to visualize the flow of water.

In this scenario, participants simulate rain on the 3D reconstructed plot using a rain particle system developed with an AFRAME physics library. This simulation will be visible on both the onsite participant's AR view as well as the remote user's VR view as a visual aid. Using this visual aid, an experienced gardener can talk a novice through planning a garden based on the flow of water on the plot (Figure 4b).

*3.3.2 Scenario 2 – Visualizing the distribution of shade.*

Participants simulate the movement of the sun over the garden by moving a spherical light source and adjust the virtual east-to-west axis in VR to align it relative to the physical plot. This can be used by an experienced gardener to highlight how the movement of the sun might affect the distribution of light on the plot over time. The webapp simulates the lighting and shadows on the 3D reconstructed plot in both the novice's AR view as well as the expert's VR view as a visual aid (Figure 4d).

The scenarios were designed to elicit feedback on existing or possible affordances for the virtual representations of areas, objects, and events in the physical garden space in XR. This includes reflecting on affordances for instructional interactions (e.g., being able to actively move a virtual sun) as well as social affordances for connecting (e.g., conveying stories for objects with family history). A limitation of using the web app approach was the constraint on the fidelity of the 3D models to maintain comparable loading and rendering times across the different mobile devices and the headset. Most models had between 10K to 100K depending on size or number of objects in the model (e.g., empty plot or single objects needed less triangles). We had one outlier of a large butterfly garden model (Figure 3f) for session six with one million triangles which we restricted to the scenario 2, since simulating rain in scenario 1 is more resource intensive due to the collision simulation for the raindrops and could negative affect participant experience with large render time.

### 3.4 Analysis

We collected 12 hours of zoom videoconference recordings of 1) onsite participants walking their remote partner and the remote researcher through the physical study locations for part 1 of the study and 2) participants using the Oculus headset or their mobile/tablet device during XR prototype evaluation in part 2 of our study. Prototype evaluation, in total, used nineteen reconstructed 3D models of the participants' garden areas and objects (Table 2). An additional 6 hours of video was screencaptured from the XR environment during prototype evaluation in part 2 that depicted participants (as avatars) interacting with each other and these 3D models.

We used a qualitative approach to analyze this rich dataset from multigenerational groups having different expertise levels and goals with gardening and perceptions of XR. We use thematic analysis (Braun and Clarke [11]) to descriptively surface the underlying subjective perspectives on context that is important for user-experience with 3D reconstructions. Our analysis was guided deductively by the study objectives, and through inductive interpretation of data. Subthemes and overarching themes related to our research questions were developed progressively using coding and memoing.

The first author first started by transcribing the audio from the videos for deeper familiarization with data. During this process, the videos screencaptured during the prototype evaluation, in part 2 of the sessions, provided context for participant reactions and actions in the "real" world videos. The first author then performed initial coding of transcripts and complementary memoing to identify emerging subthemes in data. The first four sessions (S1-S4) were transcribed and coded immediately after completion. The resulting codes and subthemes were then discussed and refined by the authoring team for their relevance to our guiding themes based on our research questions (Section 1). The first author and a research team member then transcribed the interviews from sessions S5-S8 and used this data to build on the refined subthemes. Although the codes around the research questions had become mostly saturated after the first four sessions, new data related to privacy when creating and sharing 3D content emerged during session S5. The researchers additionally focused on drawing out more data on this subtheme, when interviewing in sessions S6-S8.

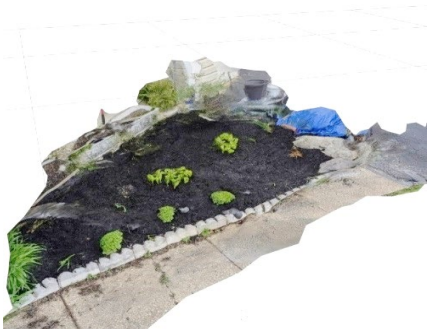

a) Decorative area in P2's frontyard
(49K triangles)

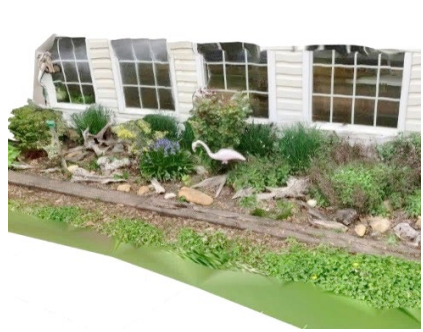

b) Decorative area in P3's farm
(80K triangles)

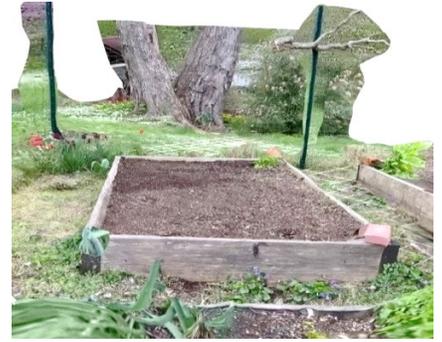

c) Garden bed in P5's backyard
(73K triangles)

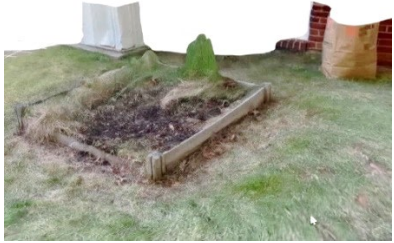

d) Garden bed in P9's backyard
(50K triangles)

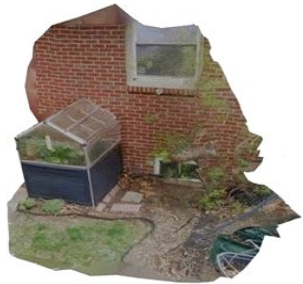

e) Herb greenhouse in P11's backyard
(106K triangles)

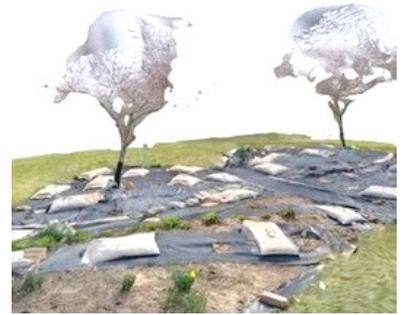

f) Butterfly garden in P12's community garden
(1.05 Million triangles)

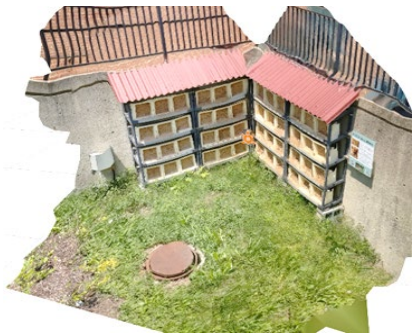

g) Bee house in P14's community garden
(83K triangles)

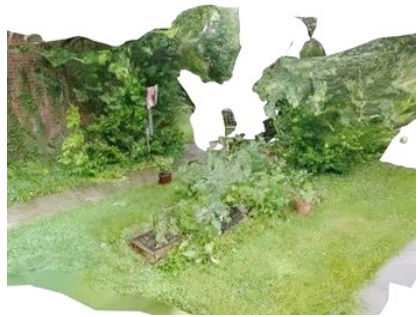

h) Garden bed in P17's front yard
(396K triangles)

Figure 3: One 3D model from the set of 3-4 models generated for each of the eight study sessions from videos clips provided by participants of garden areas and objects they selected to be useful for instruction or meaningful in some sense. The scale of these ranged between large areas (butterfly garden) and small garden beds or single objects (herb greenhouse).

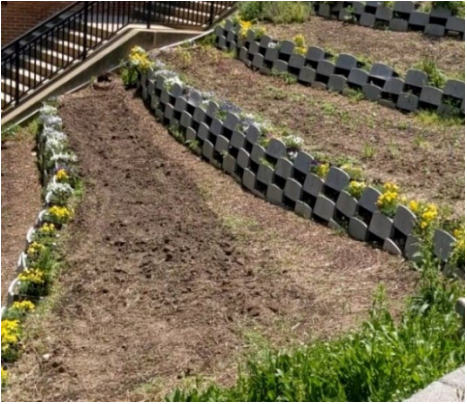 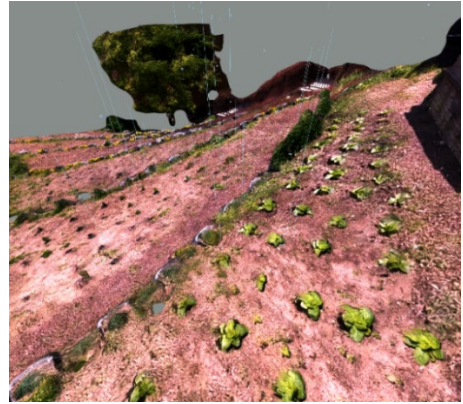

a) Still from video of terraced slope selected for session 7

b) First person view of P16 seeing rain simulated on a 3D model of the sloped terrace

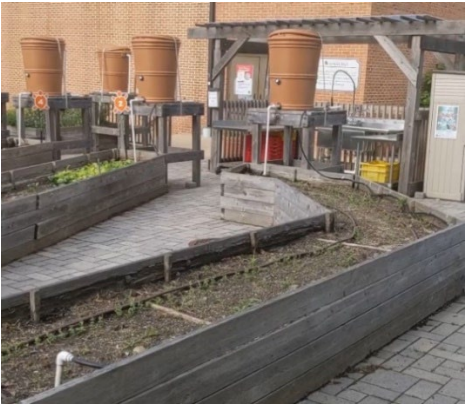 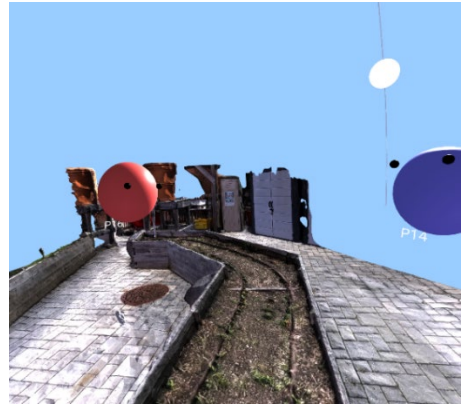

c) Still from video of raised beds and irrigation system selected for session 7

d) First person view of P15 seeing avatars of P14 and P16 experiencing scenario 2 with the virtual sun

Figure 4: Examples from session seven of areas selected by participants along with screen captures from devices used during part 2 of the study. We can see 3D models of the garden area generated and used for scenarios 1 (rain) and 2 (sun) along with the simple sphere with dotted eyes for user avatars.

The themes that emerged from coding and memos from all eight sessions were brought to our research team for further discussion to define the final salient themes that we present in the findings section. Our first salient theme relates to context that can be important to capture to reconstruct the physical activity space and objects for social connecting or instructional interactions in XR (RQ1). Our second theme relates to the goals of intergenerational groups and factors affecting their interactions with the reconstructed objects and areas in XR (RQ2). Our final theme is related to how our model generation approach might have affected prototype usage during the study sessions. The findings from this theme presented as a limitation that can influence design considerations rather than answering a specific research question. Table 3 provides examples of themes, subthemes, their associated research questions and underlying codes.

Table 3: Examples of themes, subthemes, associated research questions (related theme), and underlying codes from data analysis

| Theme | Subtheme | Research Question | Example Codes |
| --- | --- | --- | --- |
| Context important to capture for informal settings | Context for understanding the reconstruction with real-world teaching/learning experience | RQ1A | model missing surrounding visual references, viewing extremely fine details, ability to touch objects for teaching, describing an interconnected system, capturing life |
| | Context capturing relations with other people | RQ1B | Object representing family member, create familiarity with shared memory, gifting family, collaboratively built space |
| | Privacy when authoring and sharing 3D content [a] | RQ1C | Interested in sharing objects you are proud of, models of private vs public areas, privacy concerns for children |
| Intergenerational Perceptions and XR interactions | Perceptions due to values | RQ2 | Unaccustomed to internet, resonates with using technology in garden, relying on on-site people for help, younger people may be more comfortable with XR |
| | Goals for intergenerational groups | RQ2 | Teaching the old ways, helps understand where food comes from, learning from a young age can be good, shared learning between generations. |
| Effect of models on feedback | Orienting to virtual model | | Couldn't identify model as real area, resolution requirement depends on expertise, landmark helped orient |

[a] Subtheme emerged during session S5 so codes for this come from the subsequent sessions (S5-S8)

## 4 FINDINGS

In this section, we describe participants' thoughts on using the XR prototype to experience 3D models of familiar real objects and areas in their garden. We focus on what participants felt was important context that these 3D models needed to capture in XR to augment instruction and connection in their informal setting. "Context" includes behaviors, interactions, and affordances that participants would like the models to support. We also describe privacy related nuances important to participants when creating and sharing these 3D models in different settings. Finally, we describe how prototype usage may have been affected by perceptions of XR by our intergenerational groups and the models used.

### 4.1 Context important to users' experiences with the 3D reconstructed activity space in XR (RQ1)

As part of answering **RQ1**, we present our findings on contexts that cover instructional motivations **(RQ1A)** and connecting **(RQ1B)** with each other and the privacy concerns around creating and sharing such 3D content **(RQ1C)**.

*4.1.1 Context to capture for learning experiences: Exploring relations at different scales of the garden (RQ1A)*

In Figure 3, we give examples of 3D models from each session that were created using several types and scales of objects and areas chosen by participants from their physical garden space. Many smaller areas and objects in the garden can often be viewed as part of a larger natural or manmade system. This is especially visible in community gardens and demonstration gardens (S6, S7) that are designed to showcase their relationship with the surrounding local environment. For example, rainwater collected in one area of a community garden might be pumped to a different area further away to supplement existing irrigation as in S7. Master gardeners P11, P12, and P14 described how they could use models in XR to provide an overview that captured the relations between parts of the entire garden system and its interactions with the environment. Similarly, P2, P9, P11, and P14 felt 3D models could be more suited to give an overview of a garden and system rather than capturing a single object or area in isolation, if its details were recognizable. It was different when teaching/learning concepts that required the remote user to explore minute details. Some participants (P2, P5) felt video chat in the garden walkthrough allowed them to better explore these details since it "*forced you to focus on objects in a smaller field of view.*" Others emphasized creating models that emphasized zoomability (P1, P12, P18) and manipulability (P13, P14). Although limited by technology, one expected finding was for being able to manipulate the models hands-on especially in an intergenerational setting since kids "*they want hands-on, like we do leaf rubbings*".

Participants also discussed additional context when representing the effect of environmental variables on the physical activity space (sun and rain in our scenarios). An example from prototype evaluation was for the models to capture how the physical counterpart might interact with sunlight. Participants mention that this could be based on the real object's position in the house, relative geography, and its material properties when remotely assisting a user with planning. Participants felt a difference in being able to translate their real-world experience to the virtual environment in the two different scenarios for our XR prototype. With the sunlight prototypes, some participants felt they were unable to translate their experience. This was because there were many implicit cues when thinking about sunlight distribution in a real-life space that were not captured by the experience in the prototype. An example was how limit on the area for modeling affected P5 and P6's usage of the sunlight prototype. P5 indicated that the neighborhood around the object could have been better selected during video capture by including a larger area around the pre-created model to determine how sunlight interacted with her garden. However, experiencing how rainwater flowed across a model of their garden plot in the rain prototype felt more translatable and understandable by both partners and most other participants. P11 felt that being able to see where the water was draining off or feeding on the model in scenario 2 was cool since "*you know, my goal is always to conserve water as much as possible.*"

*4.1.2 Context to capture for connecting interactions: Emotions, relations with people and feelings (**RQ1B**)*

Some objects and areas in the physical space connected our participants with their partners and with other people past and present. These could be decorative like P3's neighbor's painting for their shared gardening space (Figure 2c). They could also be living and "*functional*" like P1's tobacco patch in her husband's memory, or P9's silver maple tree in memory of her mother. Some of them were also a product of collaborative effort. For example, Figure 3b) shows a model that we created of a flower bed that was pre-selected by P3 and P4 for S2. This area contains random objects like "*pieces of decorative woods*" picked up by P3 and P4's family, a gift wind catcher from P3's son, and a flamingo statue that had "*come down to live in mine*" after P3's mother passed away. Similarly, P17's house garden was full of plants (e.g., Lenten-roses) or features (e.g., winding paths) suggested by her friends and neighbors or from when "*I take walks and I see what works in other people's yards.*" Some participants wanted to create an aesthetic that appealed to other people with the objects and areas in the garden. For example, P11 wanting to create a "*tropical feel*" in his garden for his mother (P10) while also making his yard look presentable "*out of respect for my neighbors, you know*".

When discussing how one might capture these objects, areas, and associated feelings, we find some participants talking about specific kinds of temporality in addition to visual detail and interactivity that were significant. Objects in the garden also were associated with a certain time at which they held more meaning. For example, in session 1, P1 described a dogwood tree that her niece P2 had planted in memory of her mother. P2 felt that she would prefer to capture the tree as a 3D model when it was in bloom and the birds were more active. She preferred a model since she could sit under it, which would not be possible with a photograph. In addition to just seeing the actual 3D model at a certain time, P1 was personally interested in being able capture additional personalized context that would help her feel what P2 felt in some way such as P2 vocally describing the reasons for planting the tree. Similar examples appeared in sessions 5 and 7. P11 was interested in capturing a model of a memorial flower bed for his brother in bloom while P14 was thinking of modeling P16's maple from their front yard that had "*a picture of her (P16) hugging it*". Some participants describe creating and sharing models that they think would just be fun. For example, P13 felt she was a more visual person and wanted to create models of flowers that she grew that could also capture their "*vivid*" color.

### 4.1.3 Context useful for both learning and connecting: Capturing "life" and stages of growth (**RQ1A** & **RQ1B**)

Participants were also interested in context that captured a certain stage of growth or represented "*life*" in the garden space. This was valuable for instructing (**RQ1A**) and connecting (**RQ1B**) over their unsurprising shared appreciation of nature. For example, during the garden walkthrough, participants sometimes tried to familiarize their partner with how something might look when growing. After experiencing the prototype, P12 felt it would great to show something grow from seed to flowering with virtual models of plants. P18 also mentions an example of being able see the various stages of growth of a magnolia tree on a model and compare it with her own. This ties in with an interest among participants in helping others understand, like P6 for her young daughter "*where food comes from*".

Another way of "*capturing life*" in the garden was capturing the way that objects might interact with non-humans in the space and even demonstrate their common behaviors. P15 mentions being able to create a model of a certain bird that would allow others to better visualize its normal behavior and compare any odd pruning behavior in their physical counterpart. P14 described an annual event in his garden when "*tiny wren birds come in, they have babies*" and wanted to create and share a 3D model of some birdhouses that he made for this personally significant event. P18 talked about how they might want to create a 3D reconstruction of tiny praying mantis eggs hatching. She felt that experiencing this event as group with a video would involve adjusting viewing angles and "*Oh, can you come a little closer?*" whereas "*as a model, you can move more around that.*" An interesting use case suggested by P17 in the same scenario was to be able to create a model that was good enough to allow a remote user to "*look around the garden a little bit more virtually*" in real time and suggest camera positions to their onsite partner for photography. Capturing life sometimes also involved capturing models for areas that might be unapproachable, as in the case of the bee area in session S7 (Figure 3g.). In other cases, it might be that a problem of accessibility for people with mobility constraints instead. For example, P15 felt it would be nice to create a walkable model of the forest preserves adjoining her community as a way to connect through nature. It would help "*some of the are elderly shut in…(to) give them the feeling of being outside even though they're not.*"

### 4.1.4 Context when sharing 3D content: Privacy preferences related to space and activity (**RQ1C**)

An interesting finding was related to participants' privacy concerns regarding 3D content (digital twins) they might create themselves from their physical garden spaces. These concerns appeared organically from session 5 and were pursued in the following sessions as an additional research question (RQ1C). We find that participants' openness to sharing 3D representations of real spaces, either public or privately, depended on a couple of factors: how they viewed their property, the motivation for sharing, and their experience with sharing photos/videos on conventional social media. For example, P11 (younger expert) felt comfortable publicly sharing the S5 models since he viewed his garden spaces as a community resource and "*a tool to train (other people)*". He was an enthusiastic endorser of creating personalized 3D content from his own garden for educational purposes and had already created 3D CAD style models of his garden in the past. There was a similar shared opinion among a few other participants (older novice P10, older expert P12, younger expert P14, older expert P15) regarding public sharing for an educational purpose. For example, P12 was open to creating models of the educational spaces in master gardener demonstration gardens and sharing them publicly.

Participants were guarded to varying degrees when it came to sharing models from spaces viewed as private (inside of a house, backyard) or their own avatar. P10, like her son P11, did not have any concerns with publicly sharing such 3d content if the address of his house was anonymized. Some were more conscious and only wanted to share models of good-looking parts of their house, like what they might share on social media. For example, P13 was interested in creating and sharing models of her beautiful "*pet flowers*". P14 felt he would not create a model of his house garden because it was "*so bad, I don't want to show it to anybody right now*". But he also felt open to publicly or privately sharing if it "*would be a good learning tool for people who might have similar backyard or spaces.*". Some participants, like P14, felt comfortable with public sharing of models from private spaces, more so than community spaces, after scrubbing identifying information (e.g., geolocation). P15 felt she would need to ask permission when creating models in her community gardens since "*I wouldn't feel right doing that. Because it's not my property, there are common areas and there are other households there.*" These data points highlighted contrasting views that can exist regarding the extent of information that needed to be anonymized. Another example comes from S8 where P17 was not on social media and P18 only occasionally posted pictures of her flowers for her friends on Instagram, but nothing too personal. P18 did not feel hung up about even having her house number visible. To a random viewer "*this is the number of a house in who knows where ville*". P17 on the other hand felt someone could "*match it on google earth*". Another key point that P12 raised was regarding self-representation. Our XR prototype avatars felt "*really basic*" but the ability to create a lookalike avatar of oneself posed privacy concerns and required guidance when children were involved in the intergenerational setting. P15 (older adult) and P16 (teen) on the other hand felt comfortable with being recognized as themselves.

### 4.2 Interplay between perception of XR and prototype usage in our intergenerational groups (RQ2)

An observation that we would like to highlight from our study procedures is the interplay between learning to work with the prototypes together and the multigenerational perceptions of XR. Most of our participants were aware of the term virtual reality but every session had atleast one participant who was experiencing VR hands-on for the first time (Table 1). On one hand, participants had positive inputs regarding creating and using digital twins of physical spaces and objects for the purpose of instruction. Not just "*for games but for actually, you know, something more purposeful and useful such as all the benefits of gardening, for example, physical emotional mental, and community, and also environmental*" (P15). Participants were interested in how such technology could be used for sharing of knowledge such as passing on "*the old ways*" of tradition (P3) or educating older generations on modern sustainability techniques (P11). Others were interested in connecting by listening to family stories from their garden space (P9). However, it is also important to acknowledge that there were participants who did not see themselves using XR devices in the garden. This was in spite of liking the idea of capturing different objects or spaces and the related contexts as 3D models to view using XR. Some participants, such as P1 (older novice), P2 (younger expert), P3 (older expert), P4(experienced teen gardener) talk about how not using technology in the garden resonated with their values when they were living close by. They felt it was probably better suited for the presented use case of families or friends separated by distance when there "*was no way to have personal interaction*". P4 talks about enjoying nature as "*a lot of this [her] generation is involved with a lot of technology, so I think getting out in the garden, you know, removes you from [technology]*".

P3 (older expert), P7(older expert), P8(older expert) felt that experiencing a remote garden using VR technology might appeal to people who were "*much more familiar with the computer*" (P7) or who "*really like technology*" (P8). While they do acknowledge that there are other older adults who fit this category, they also mention that "*the younger generation*" (P8) would feel much more comfortable using such technology. P3 even mentions that young kids like her granddaughter P4 might find it more interesting "*as you can see, P4 was just fascinated with what she was doing while she had [the headset]*". These specific older adult participants however seemed willing to try such a prototype because it was an intergenerational setting with someone they knew. We find that it may have helped for participants to have assistance from their older or younger study partner, or just bounce ideas off them when troubleshooting issues with the device or prototype (e.g., orienting to the model of the garden based on familiarity). Even P10 (older novice) who was an academic and familiar with new technologies mentioned being excited to try the technology and having P11 walk her through the using the oculus quest headset before the session. We have previously already noted the values of our participants related to technology in the garden and how XR might be viewed as unnecessary by some regardless of age or tech familiarity. The motivation to share knowledge and the dynamic of helping each other with using the prototype may have enabled us to obtain more positive feedback than expected in spite of these perceptions of XR for a garden.

### 4.3 Effect of model quality on prototype feedback as a limitation

The video collection method to generate models using photogrammetry might have been a factor that affected prototype usage and feedback. As mentioned earlier in section 3.3, the amount data used in the photogrammetry process and the resulting fidelity of the models presented to our participants was limited by the web app approach for our prototype. For this reason, some models initially felt unfamiliar to participants. They required time to orient to major landmarks and view the model as natural. P3 felt the model of her farm plot resembled more of a "*Martian landscape*" and the barnyard acted as a unique landmark (Figure 5a) to help their group orient themselves. A functional use of the models, for example, when simulating rain in scenario 2 also helped participants overlook the model quality issues. Some, like P12 and P13, in S6 were impressed with the quality of the models we were able to create for them, even after initially feeling lost. P12 mentions that people, might have different opinions about how "*vivid or lifelike*" they might want the model they capture to be for instructional purposes, depending on their level of expertise.

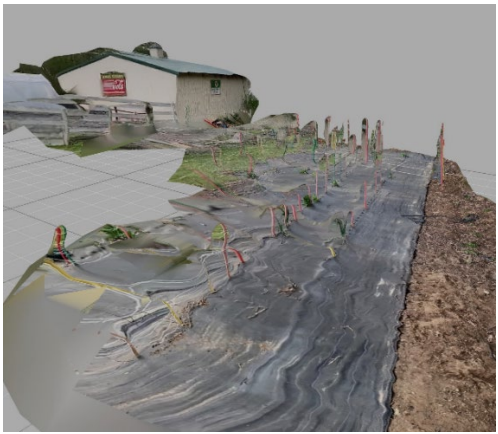
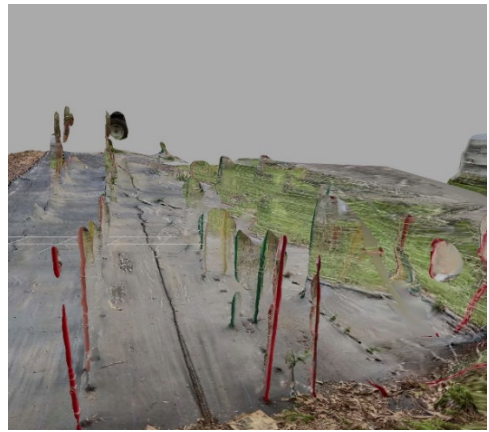

a) Viewpoint for model of P3's farm plot which was relatively recognizable with the large barn landmark. Objects like the vertical stakes look distorted.

b) Alternate viewpoint of the model showing issues with directional fidelity. Some of the vertical stakes look better here

Figure 5: One of the models for session 2 which participants described as difficult to recognize and unnatural like a *"Martian landscape"*. Corresponding photograph of the physical area in Figure 2b.

The fidelity of the models also was directional, with the XR users seeing better visual quality with certain viewing angles and distortion of the models from other angles (Figure 5b). This might have been due to the technique suggested to participants to collect videos for model creation by moving around an area or object which resulted in more photos from angles where the users focused on objects longer. The viewing angles explored in the videos might also have been influenced by the movement area available to the participants since gardens are often full of objects that restrict mobility outside of a built path (e.g., bushes, rocks, muddy terrain). While the videos for photogrammetry were captured from a distance that allowed the entire area or object to be visible, there was a variation in this distance depending on the target's size. This might have caused variation in capturing the neighborhood of the target object or area which was a key factor in realistically visualizing environmental effects, as mentioned earlier in Section 4.2.1 by P5. Overall, the low-fidelity constraints and visual artifacts in the 3D models might have affected their perception of the state of XR and how one might interact with these models.

## 5 DISCUSSION

Our findings describe context from informal settings that participants felt was important to user experience when interacting with virtual reconstructions of real areas and objects in XR. In this section, we discuss what these findings might mean for perspectives on creating and sharing 3D models of meaningful areas and objects. We focus on how they fit into the space of 3D contextual capture for instruction and connection.

### 5.1 Capturing context for meaningful reconstructions of objects and spaces for informal XR learning

First, we discuss possible limitations for XR user experience and approaches to capturing context through 3D models when instruction involves objects at different scales. We also note potential challenges with translating real-world experience to reconstruction of objects and spaces.

*5.1.1 Supporting meaningful reflection during remote instruction (**RQ1A**)*

Our findings described how the scale of the selected meaningful objects and spaces allowed for discussion related to different objectives. Video seemed more suitable when remote learning involved focusing on minute details, while XR seemed more suitable for visualizing relatively larger systems and their relationship with the environment. Past work around multi-scale interactions in XR, has also noted lower visual quality as a tradeoff for exploration with 3D models over video in other remote-guidance settings [55,80]. So, while video might be sufficient to capture context at smaller scales, designers might benefit from leaning more into enabling the user to change perspective at different scales with XR. This can augment different kinds of reflection or instruction in remote collaboration systems (e.g. first person and third person [32,34,76] , giant-miniature [55], using 360 video [30,77,79] or 3d reconstruction [82]).

    User experience with interactions at these different scales and perspectives could benefit from exploring different approaches to reconstruction. This could help identify other types of contexts to capture for meaningful reconstruction which could augment reflective and shared memory practices among the users about the activity and the physical space and objects. Our findings noted that the approach to capturing the models and reconstruction artifacts posed some limitations. A physical constraint of the data collection method we used for reconstruction was that participants had to select objects and spaces that were easier to walk around. Consequently, objects needed to have features that were visible from the walked path. The models did allow the participants to reflect on certain types of contexts to capture at this scale like socially shareable features, interaction with rain, or animals. A hands-on physical approach to learning, as in gardening, might be better served through a technique like in-hand object reconstruction [88] from first person video. This might be useful for collaborative exploration of minute details by naturally selecting and pointing at important context for hand-scale objects with hand-interactions. The interactions in this approach could possibly reveal affordances (e.g., tactile) that onsite users would want the model to have as context for the remote user.

Again, conventional video that is zoomed in might be more preferred for minute details and as such could be a benchmark when comparing more interactive ways of reconstructing objects with fine details for remote collaboration.

We have also seen participants discussing the advantages of seeing the relations between different objects and areas as part of a garden system and the surrounding environmental context. Enabling a learner to visualize a particular system component within a broader landscape can augment how they reflect on its role. YardMap is one such example project, where crowdsourced labelling of 2D neighborhood maps allows for reflection on land use and wildlife behaviors [94]. 3D representations in XR can be a powerful perspective-taking medium to reflect on environmental effects of human activity[18]. Researchers could explore how individual, or group input could help reflect on and identify context for reconstruction that can support reflection for people of all expertise levels. For example, by deciding a meaningfully sized neighborhood to gauge the impact of environmental variables like rain. Designing reconstruction for this purpose could involve allowing users to capture context surrounding a specific object or view point [35], for example, from a bird's-eye drone or third-person perspective [67], or 360 imagery.

### 5.1.2 More implicit context required to understand 3D models with real-world experience (**RQ1A** and **RQ1B**)

"*If a picture is worth a thousand words, one could argue that a model is worth a thousand pictures, or a million words.* [84]" However, this depends on whether there is contextualizing metadata to help with assessing and reusing the model. Past work has explored adding different contextual information (e.g. user annotations, sensor data, audio) to conventional 2D media to enable different interactions for sensemaking and creative expression [3,23,24]. Our findings show examples of such context that participants felt were important when providing instruction or for social connection using 3D reconstructions. Providing a shared visual context explorable in an embodied manner is an advantage of XR systems for remote collaboration [10,72]. However, our findings (Section 4.1.1) show that sometimes scenarios that are assumed to be intuitive for simulation with a reconstructed model aren't necessarily so. There is implicit or intangible context [89] from real world experience that is lost even if we have a high-fidelity reconstruction of a 3D environment. We find this in the case of sunlight visualization over time using the reconstructions. We had selected it as a promising use case for visualizing time in XR from Maddali et. al's [40] study on use-cases for expert novice with 360 image environments. However, the feedback on this scenario shows that it required much more implicit context, for example, a meaningfully sized neighborhood of influence around the object.

Our findings also talk about themes such as capturing "life" and growth that might be expected from a nature setting. These were operationalized for reconstruction by participants as being able to show visuals at various stages in time, or as behaviors expected from objects or non-human beings that inhabit the space. The idea was to have a reference to compared with a physical counterpart and evaluate, for example, growth or normal behavior. Being able to virtually represent animals and their presence (e.g. ambient sound of environment) or behavior (e.g. swarming) authentically, especially from an environment inaccessible to people, can augment connecting interactions in environmental education [40,78]. However, reconstruction of animals is an especially challenging task and doesn't lend itself well to live settings, since they aren't static and rigid objects [4]. This limits how personizable such reconstructions can be for informal connecting interactions compared to traditional videos/images. As a design consideration, rather than the ability to create high fidelity models, searchability and sharing of models might be more important depending on whether the objective of the users is learning or connecting. Digital sharing approaches could just make existing libraries of professionally constructed virtual 3D models easily accessible on the interface for social interaction (e.g. AR animals in google searches [74]) or even searchable (e.g. for a young versus grown flowering plant).

## 5.2 Reflecting on privacy nuances when creating or sharing 3D models for XR (RQ1C)

The growing ubiquity of XR for casual use adds yet another set of devices for which users must reflect on personal privacy implications. XR devices pose a unique challenge since they inherently rely on mapping their surrounding space to connect the physical surrounding with their virtual environments and enable spatial interactions [43]. Industry initiatives (e.g. Niantic's Planet-Scale AR Alliance [106]) envision a future where "world scraping" [29] through XR devices will allow for "live maps" [98] of reconstructed 3D spaces to the level of specific object locations. This kind of mapping can enable useful applications, for example, accessible navigation [99]. However, they also further blur the idea of what XR users consider as private space or objects and what context is private specifically for reconstructions in XR.

Our findings describe a few perspectives on privacy when creating or sharing 3d reconstructions as a casual XR usual. The process of capturing and sharing models could be influenced by participants' thoughts on what was a private or public space and what context they might want to share. Wang et al. [85] describe this context for the casual sharing of dance movements as 3D reconstructed moments. They also acknowledge that privacy considerations might change with in different contexts as informed by previous work by Li et. al on photo-sharing [37]. Our findings confirm this assumption and extend past ethnographic findings on sociality in such informal spaces. We have provided an example for how educational motivations were more correlated with openness to public sharing of 3D models from personal garden spaces. Users should be encouraged, through design, to define privacy for spaces, objects, and any important related context. An approach from past work is to help casual users reflect and identify context whose importance to privacy might be overlooked during reconstruction [81,87]. Designing for visibility of motivations and preferences related to object considered private could be another approach to inform outsiders reconstructing or sharing a model of the same object [41]. For example, allowing the user to select a space around an object to provides enough context to visualize interaction with the sun but hides a house number or a neighbor's plot. Further, in an intergenerational setting, it is imperative that designers balance protecting the privacy of more vulnerable members of the activity space with capturing meaningful context when creating and sharing models [107]. Our findings provide an example of the possible risk of self-representation using a reconstructed avatar that younger users might not recognize. Future work could specifically try to understand how these vulnerable members might approach the context capture process to better tailor it for their privacy preferences.

## 5.3 Capturing context from the garden for social connection and intergenerational XR

Our findings provide an understanding of context that can capture social relations and emotional context. We discuss implications for authoring and sharing 3d content that can represent shared memories. We also discuss settings outside of gardens where exploring meaningful reconstruction of spaces and objects could augment intergenerational learning and social interactions.

### 5.3.1 Value of reconstruction in preserving shared memories (RQ1B)

Personal environments, like the home, are often likened to an "autotopography", a spatial and physical representation of their people's individual story [21,54] as we've seen in our own findings. Shewbridge et. al in a cultural probe study discuss how, for a future where 3D printers are integrated into a home environment [69], participants often selected unique artifacts in their house with sentimental value as something they would want to scan and create accurate or similar physical replicas. We also find ideas in literature on using immersive media (e.g. 360 panoramas) to preserve the memory and feelings associated with notable individuals' private spaces by reconstructing how their belongings were arranged [35]. Only recently, there has also been commercial interest in making it easier for everyday users to create 3D or 4D reconstructions of personal environments or shared moments using photos and videos [96,108].

Community and hobby spaces like gardens can be viewed by researchers as a sociotopography in addition to an autotopography. Here, objects and spaces are ascribed a shared meaning by groups (e.g. family, or practitioners) [41]. Our findings present examples of these and some shared motivations for why participants might want to have virtual representations they would like to use in remote interactions. We identify a few related design considerations from Section 4.1.2, where participants wanted to reconstruct an object or space representing a generational connection. First, it was important to identify context that represents the object a particular time (e.g., the time of bloom). Second, there can be some embodied interactions with the virtual representations that can be meaningful to user experience: to hug it. For example, sitting under a virtual replica of a tree can capture context that represents nostalgia using XR. Finally, there might be a need to contextualize the sensory reconstruction with other social data, like a narrative [21] from the person who finds it meaningful. This can be helpful for a viewer to make sense of the same context.

There is a growing interest among HCI researchers in understanding the role of digital technology in shared memory practices through creating and sharing artefacts. We encourage researchers to think about whether the virtual reconstructions can augment preserving family memories and heritage on a more personal level as part of casual XR, in a similar way that reconstructions of museum artifacts preserving cultural heritage [35]. A question to consider is how exactly these reconstructions would be perceived in terms of value for reminiscence in intergenerational settings or as mementos that are useful for connecting interactions. There is much work on comparing physical with digital mementoes, where the physical mementoes arguably requiring more narrative to explain their significance and digital portrayed as less emotionally expressive [53]. Would virtual reconstructions in XR be valued closer to conventional digital artifacts, like 2D photos or videos since they inherently have no physicality without being 3D printed or viewed through a XR device? Alternatively, virtual reconstructions could also be seen as closer to physical artifacts since our participants felt they can be experienced in an embodied manner in XR. This can have interesting implications for how memories are captured, shared, and used in connecting interactions in the future.

*5.3.2 Other physical spaces for casual XR as a medium for intergenerational activities (**RQ2**)*

If we imagine the future of remote social interaction to be tied in some way to XR, we will ideally want XR to be welcoming and accessible [97] to users of different ages, abilities, and backgrounds while respecting their motivations, values for activities chosen as use cases, and privacy considerations. Our findings in section 4.2 presented some perceptions of whether intergenerational groups would use XR for such remote collaboration in the garden. We see that these perceptions were not necessarily related to age and might have been more correlated with self-described tech familiarity. The values that arose from gardens being a nature experience removed from technology, as described in past literature [5,22,40], also influenced them. Most of their positive experiences around interacting with the 3D reconstructions were centered around sharing knowledge between generations, encouraging reflection (e.g., on environmental effect), and representing shared memories. These are not exclusive to the garden space.

Past work has studied learning spaces in other activities notably characterized by intergenerational participation. This includes more human-made-artifact oriented activity spaces such as makerspaces [52] where virtual representations of objects using technology have become integrated into the process of ideation and making [83,86]. Artisanal activities entrenched in tradition (e.g. pottery, papermaking) have also seen interest from HCI researchers as settings for capturing context representing tacit knowledge and augmenting learning using XR [27]. The applicability of our findings on experience with 3D reconstructions of real spaces and objects in these activities is a question that might interest designers. While hobbyist values might be comparable across activities on a high level, preferences for technology might differ [22]. A factor for this could be that technology might blend in better with, for example a makerspace environment compared to a natural space as we have seen in our findings. Other relevant, more everyday activity spaces might be found inside the home. such as the kitchen. Panicker et. al [50] encourage supporting and creating shared experiences around eating and meal preparation by, for example, facilitating cooking together across distances. The XR industry is in the process of finding more of these meaningful and practical everyday use cases for casual users. At the risk of sounding techno-solutionist, we would like to encourage more support for exploring applications that can meaningfully augment remote group experiences for an intergenerational audience. There is clearly an interest in the potential applications for remote XR to go beyond entertainment and work productivity, but the perception of being an accessible and general-purpose remote collaboration device is not there yet.

Table 4: Salient Findings and Implications

| Research Question | Findings | Implications |
|---|---|---|
| **RQ1A** | • XR better for learning about larger systems in garden compared to video for minute features.<br>• Implicit context not easy to understand from 3D models<br>• Interest in capturing "life"-like behavior and comparing with real. | • Capture context for perspective taking and reflection using reconstruction at different scales<br>• Allow users to define a meaningful neighborhood around object to better capture implicit context |
| **RQ1B** | • Time or state of object can be meaningful (e.g., in bloom)<br>• Embodied experience with models is meaningful (e.g., sit under)<br>• Narrative context helpful to understand relation with model. | Ambiguity in perception of value of 3d representations vs 2D media. Context, like embodied interaction, is a plus for XR |
| **RQ1C** | • Educational motivations positively correlate with sharing 3D models of personal spaces<br>• Contrasting views on extent of context (e.g., location) to anonymize. | • Design to encourage user to reflect and define privacy for space and meaningful context.<br>• Designing for visibility of motivations and preferences related to objects to outsiders. |
| **RQ2** | • Contrasting views on using 3D models vs onsite in nature.<br>• Perception of appealing to younger or more tech-familiar users<br>• Positive intergenerational dynamic when exploring XR ideas using the prototype | • Learning & remote family use-cases still meaningful.<br>• Contrasting experiences from a tech-integrated activity space (e.g., indoor) suggested. |

## 6 CONCLUSION

Through an XR prototype evaluation study with eight intergenerational groups of 18 closely related gardeners, we provide an understanding of the context required when creating meaningful virtual reconstructions of physical spaces and objects to enable instructional and connecting interactions in informal settings for XR users. Participants linked health, creative expression, and intergenerational knowledge sharing motivations with objects and areas in their gardens holding shared meaning. We find that these motivations translated to reconstruction requiring context that captured relations between areas and objects at different scales, emotional context for relations with other people, and privacy considerations when creating or sharing the 3d models. We discuss implications for user involvement to create reconstructions that better translate real-world experience, encourage reflection, incorporate privacy considerations, and preserve shared experiences with XR as a medium for informal intergenerational activities.

## ACKNOWLEDGEMENTS

This work was supported, in part, by grant 90REGE0008, U.S. Admin. for Community Living, NIDILRR, Dept. of Health and Human Services. Opinions expressed do not necessarily represent official policy of the Federal government. We thank the members of the University of Maryland and its Extension Master Gardener Program for their participation in this project.

## REFERENCES

[1] Karan Ahuja, Deval Shah, Sujeath Pareddy, Franceska Xhakaj, Amy Ogan, Yuvraj Agarwal, and Chris Harrison. 2021. Classroom Digital Twins with Instrumentation-Free Gaze Tracking. In *Proceedings of the 2021 CHI Conference on Human Factors in Computing Systems* (CHI '21), Association for Computing Machinery, New York, NY, USA, 1–9. DOI:https://doi.org/10.1145/3411764.3445711

[2] Sultan A. Alharthi, Katta Spiel, William A. Hamilton, Elizabeth Bonsignore, and Zachary O. Toups. 2018. Collaborative Mixed Reality Games. In *Companion of the 2018 ACM Conference on Computer Supported Cooperative Work and Social Computing* (CSCW '18), Association for Computing Machinery, New York, NY, USA, 447–454. DOI:https://doi.org/10.1145/3272973.3273013

[3] Chris Baber, James Cross, Tariq Khaleel, and Russell Beale. 2008. Location-based photography as sense-making. In *Proceedings of the 22nd British HCI Group Annual Conference on People and Computers: Culture, Creativity, Interaction - Volume 1* (BCS-HCI '08), BCS Learning & Development Ltd., Swindon, GBR, 133–140.

[4] Marc Badger, Yufu Wang, Adarsh Modh, Ammon Perkes, Nikos Kolotouros, Bernd G. Pfrommer, Marc F. Schmidt, and Kostas Daniilidis. 2020. 3D Bird Reconstruction: A Dataset, Model, and Shape Recovery from a Single View. In *Computer Vision – ECCV 2020: 16th European Conference, Glasgow, UK, August 23–28, 2020, Proceedings, Part XVIII*, Springer-Verlag, Berlin, Heidelberg, 1–17. DOI:https://doi.org/10.1007/978-3-030-58523-5_1

[5] Eric P.S. Baumer and M. Six Silberman. 2011. When the Implication is Not to Design (Technology). In *Proceedings of the SIGCHI Conference on Human Factors in Computing Systems* (CHI '11), ACM, New York, NY, USA, 2271–2274. DOI:https://doi.org/10.1145/1978942.1979275

[6] Pieter J. Beers, Henny P. A. Boshuizen, Paul A. Kirschner, and Wim H. Gijselaers. 2006. Common Ground, Complex Problems and Decision Making. *Group Decis Negot* 15, 6 (November 2006), 529–556. DOI:https://doi.org/10.1007/s10726-006-9030-1

[7] Peter Bennett, Mike Fraser, and Madeline Balaam. 2012. ChronoTape: tangible timelines for family history. In *Proceedings of the Sixth International Conference on Tangible, Embedded and Embodied Interaction* (TEI '12), Association for Computing Machinery, New York, NY, USA, 49–56. DOI:https://doi.org/10.1145/2148131.2148144

[8] Peter Bennett, Heidi Hinder, Seana Kozar, Christopher Bowdler, Elaine Massung, Tim Cole, Helen Manchester, and Kirsten Cater. 2015. TopoTiles: Storytelling in Care Homes with Topographic Tangibles. In *Proceedings of the 33rd Annual ACM Conference Extended Abstracts on Human Factors in Computing Systems* (CHI EA '15), Association for Computing Machinery, New York, NY, USA, 911–916. DOI:https://doi.org/10.1145/2702613.2732918

[9] George R. Bent, David Pfaff, Mackenzie Brooks, Roxanne Radpour, and John Delaney. 2022. A practical workflow for the 3D reconstruction of complex historic sites and their decorative interiors: Florence As It Was and the church of Orsanmichele. *Heritage Science* 10, 1 (July 2022), 118. DOI:https://doi.org/10.1186/s40494-022-00750-1

[10] Mark Billinghurst and Hirokazu Kato. 2002. Collaborative Augmented Reality. *Commun. ACM* 45, 7 (July 2002), 64–70. DOI:https://doi.org/10.1145/514236.514265

[11] Virginia Braun and Victoria Clarke. 2006. Using Thematic Analysis in Psychology. *Qualitative research in psychology* 3, 2 (2006), 77–101.

[12] Margot Brereton, Alessandro Soro, Kate Vaisutis, and Paul Roe. 2015. The Messaging Kettle: Prototyping Connection over a Distance Between Adult Children and Older Parents. In *Proceedings of the 33rd Annual ACM Conference on Human Factors in Computing Systems* (CHI '15), ACM, New York, NY, USA, 713–716. DOI:https://doi.org/10.1145/2702123.2702462

[13] Rohan Chabra, Adrian Ilie, Nicholas Rewkowski, Young-Woon Cha, and Henry Fuchs. 2017. Optimizing placement of commodity depth cameras for known 3D dynamic scene capture. In *2017 IEEE Virtual Reality (VR)*, 157–166. DOI:https://doi.org/10.1109/VR.2017.7892243

[14] Wen-Huei Chou, Yi-Chun Li, Ya-Fang Chen, Mieko Ohsuga, and Tsuyoshi Inoue. 2022. Empirical Study of Virtual Reality to Promote Intergenerational Communication: Taiwan Traditional Glove Puppetry as Example. *Sustainability* 14, 6 (January 2022), 3213. DOI:https://doi.org/10.3390/su14063213

[15] Herbert H. Clark (Ed.). 1996. Common ground. In *Using Language*. Cambridge University Press, Cambridge, 92–122. DOI:https://doi.org/10.1017/CBO9780511620539.005

[16] Sheri T. Dorn, Milton G. Newberry, Ellen M. Bauske, and Svoboda V. Pennisi. 2018. Extension Master Gardener Volunteers of the 21st Century: Educated, Prosperous, and Committed. *HortTechnology* 28, 2 (April 2018), 218–229. DOI:https://doi.org/10.21273/HORTTECH03998-18

[17] Carmine Elvezio, Mengu Sukan, Ohan Oda, Steven Feiner, and Barbara Tversky. 2017. Remote collaboration in AR and VR using virtual replicas. In *ACM SIGGRAPH 2017 VR Village* (SIGGRAPH '17), Association for Computing Machinery, New York, NY, USA, 1–2. DOI:https://doi.org/10.1145/3089269.3089281

[18] Geraldine Fauville, Anna C. M. Queiroz, Linda Hambrick, Bryan A. Brown, and Jeremy N. Bailenson. 2020. Participatory research on using virtual reality to teach ocean acidification: a study in the marine education community. *Environmental Education Research* 0, 0 (2020), 1–25. DOI:https://doi.org/10.1080/13504622.2020.1803797

[19] Lei Gao, Huidong Bai, Weiping He, Mark Billinghurst, and Robert W. Lindeman. 2018. Real-time visual representations for mobile mixed reality remote collaboration. In *SIGGRAPH Asia 2018 Virtual & Augmented Reality* (SA '18), Association for Computing Machinery, New York, NY, USA, 1–2. DOI:https://doi.org/10.1145/3275495.3275515

[20] Danilo Gasques, Janet G. Johnson, Tommy Sharkey, Yuanyuan Feng, Ru Wang, Zhuoqun Robin Xu, Enrique Zavala, Yifei Zhang, Wanze Xie, Xinming Zhang, Konrad Davis, Michael Yip, and Nadir Weibel. 2021. ARTEMIS: A Collaborative Mixed-Reality System for Immersive Surgical Telementoring. In *Proceedings of the 2021 CHI Conference on Human Factors in Computing Systems.* Association for Computing Machinery, New York, NY, USA, 1–14. Retrieved October 24, 2021 from https://doi.org/10.1145/3411764.3445576

[21] Jennifer A. Gonzalez. 1995. Autotopographies. *Prosthetic territories: Politics and …* (January 1995). Retrieved August 25, 2022 from https://www.academia.edu/2217199/Autotopographies

[22] Elizabeth Goodman and Daniela Rosner. 2011. From Garments to Gardens: Negotiating Material Relationships Online and "by Hand." In *Proceedings of the SIGCHI Conference on Human Factors in Computing Systems* (CHI '11), ACM, New York, NY, USA, 2257–2266. DOI:https://doi.org/10.1145/1978942.1979273

[23] Maria Håkansson and Lalya Gaye. 2008. Bringing context to the foreground: designing for creative engagement in a novel still camera application. In *Proceedings of the 7th ACM conference on Designing interactive systems* (DIS '08), Association for Computing Machinery, New York, NY, USA, 164–173. DOI:https://doi.org/10.1145/1394445.1394463

[24] Maria Håkansson, Sara Ljungblad, Lalya Gaye, and Lars Erik Holmquist. 2006. Snapshots from a study of context photography. In *CHI '06 Extended Abstracts on Human Factors in Computing Systems* (CHI EA '06), Association for Computing Machinery, New York, NY, USA, 333–338. DOI:https://doi.org/10.1145/1125451.1125525

[25] Jennifer Healey, Duotun Wang, Curtis Wigington, Tong Sun, and Huaishu Peng. 2021. A Mixed-Reality System to Promote Child Engagement in Remote Intergenerational Storytelling. In *2021 IEEE International Symposium on Mixed and Augmented Reality Adjunct (ISMAR-Adjunct)*, 274–279. DOI:https://doi.org/10.1109/ISMAR-Adjunct54149.2021.00063

[26] Sara Heitlinger, Nick Bryan-Kinns, and Janis Jefferies. 2013. Sustainable HCI for Grassroots Urban Food-Growing Communities. In *Proceedings of the 25th Australian Computer-Human Interaction Conference: Augmentation, Application, Innovation, Collaboration* (OzCHI '13), Association for Computing Machinery, New York, NY, USA, 255–264. DOI:https://doi.org/10.1145/2541016.2541023

[27] Atsushi Hiyama, Hiroyuki Onimaru, Mariko Miyashita, Eikan Ebuchi, Masazumi Seki, and Michitaka Hirose. 2013. Augmented Reality System for Measuring and Learning Tacit Artisan Skills. In *Human Interface and the Management of Information. Information and Interaction for Health, Safety, Mobility and Complex Environments*, Springer Berlin Heidelberg, Berlin, Heidelberg, 85–91.

[28] H.G. Hoffman. 1998. Physically touching virtual objects using tactile augmentation enhances the realism of virtual environments. In *Proceedings. IEEE 1998 Virtual Reality Annual International Symposium (Cat. No.98CB36180)*, 59–63. DOI:https://doi.org/10.1109/VRAIS.1998.658423

[29] Adrian Hon. 2020. Digital Sight Management, and the Mystery of the Missing Amazon Receipts. Retrieved August 30, 2022 from https://mssv.net/2020/08/16/digital-sight-management-and-the-mystery-of-the-missing-amazon-receipts/

[30] Kevin Huang, Jiannan Li, Mauricio Sousa, and Tovi Grossman. 2022. immersivePOV: Filming How-To Videos with a Head-Mounted 360° Action Camera. In *Proceedings of the 2022 CHI Conference on Human Factors in Computing Systems* (CHI '22), Association for Computing Machinery, New York, NY, USA, 1–13. DOI:https://doi.org/10.1145/3491102.3517468

[31] Lisa Brown Jaloza. 2019. Inside Facebook Reality Labs: Research updates and the future of social connection. *Tech at Meta*. Retrieved August 24, 2022 from https://tech.fb.com/ar-vr/2019/09/inside-facebook-reality-labs-research-updates-and-the-future-of-social-connection/

[32] Shunichi Kasahara and Jun Rekimoto. 2015. JackIn head: immersive visual telepresence system with omnidirectional wearable camera for remote collaboration. In *Proceedings of the 21st ACM Symposium on Virtual Reality Software and Technology* (VRST '15), Association for Computing Machinery, New York, NY, USA, 217–225. DOI:https://doi.org/10.1145/2821592.2821608

[33] Eng Tat Khoo, Tim Merritt, and Adrian David Cheok. 2009. Designing physical and social intergenerational family entertainment. *Interacting with Computers* 21, 1–2 (January 2009), 76–87. DOI:https://doi.org/10.1016/j.intcom.2008.10.009

[34] Ryohei Komiyama, Takashi Miyaki, and Jun Rekimoto. 2017. JackIn Space: Designing a Seamless Transition Between First and Third Person View for Effective Telepresence Collaborations. In *Proceedings of the 8th Augmented Human International Conference* (AH '17), ACM, New York, NY, USA, 14:1-14:9. DOI:https://doi.org/10.1145/3041164.3041183

[35] Karol Kwiatek. 2012. How to preserve inspirational environments that once surrounded a poet? Immersive 360° video and the cultural memory of Charles Causley's poetry. In *2012 18th International Conference on Virtual Systems and Multimedia*, 243–250. DOI:https://doi.org/10.1109/VSMM.2012.6365931

[36] Cun Li, Jun Hu, Bart Hengeveld, and Caroline Hummels. 2019. Story-Me: Design of a System to Support Intergenerational Storytelling and Preservation for Older Adults. In *Companion Publication of the 2019 on Designing Interactive Systems Conference 2019 Companion* (DIS '19 Companion), Association for Computing Machinery, New York, NY, USA, 245–250. DOI:https://doi.org/10.1145/3301019.3323902

[37] Yifang Li, Nishant Vishwamitra, Hongxin Hu, and Kelly Caine. 2020. Towards A Taxonomy of Content Sensitivity and Sharing Preferences for Photos. In *Proceedings of the 2020 CHI Conference on Human Factors in Computing Systems* (CHI '20), Association for Computing Machinery, New York, NY, USA, 1–14. DOI:https://doi.org/10.1145/3313831.3376498

[38] Stephen Lombardi, Jason Saragih, Tomas Simon, and Yaser Sheikh. 2018. Deep appearance models for face rendering. *ACM Trans. Graph.* 37, 4 (July 2018), 68:1-68:13. DOI:https://doi.org/10.1145/3197517.3201401

[39] Peter Lyle, Jaz Hee-jeong Choi, and Marcus Foth. 2015. Growing Food in the City: Design Ideations for Urban Residential Gardeners. In *Proceedings of the 7th International Conference on Communities and Technologies* (C&T '15), ACM, New York, NY, USA, 89–97. DOI:https://doi.org/10.1145/2768545.2768549

[40] Hanuma Teja Maddali, Andrew Irlitti, and Amanda Lazar. 2022. Probing the Potential of Extended Reality to Connect Experts and Novices in the Garden. In *ACM Hum.-Comput. Interact*, 29. DOI:https://doi.org/10.1145/3555211

[41] Hanuma Teja Maddali and Amanda Lazar. 2020. Sociality and Skill Sharing in the Garden. In *Proceedings of the 2020 CHI Conference on Human Factors in Computing Systems* (CHI '20), Association for Computing Machinery, New York, NY, USA, 1–13. DOI:https://doi.org/10.1145/3313831.3376246

[42] Thomas M. Malaby. 2007. Beyond Play. *Games and Culture* 2, 2 (2007), 95. Retrieved September 2, 2022 from https://www.academia.edu/239795/Beyond_Play_A_New_Approach_to_Games

[43] McGill and Mark. 2021. The IEEE Global Initiative on Ethics of Extended Reality (XR) Report–Extended Reality (XR) and the Erosion of Anonymity and Privacy. *Extended Reality (XR) and the Erosion of Anonymity and Privacy - White Paper* (November 2021), 1–24.

[44] Vibhav Nanda, Hanuma Teja Maddali, and Amanda Lazar. 2022. Does XR introduce experience asymmetry in an intergenerational setting? ACM, Athens, Greece. DOI:https://doi.org/10.1145/3517428.3550373

[45] Ohan Oda, Carmine Elvezio, Mengu Sukan, Steven Feiner, and Barbara Tversky. 2015. Virtual Replicas for Remote Assistance in Virtual and Augmented Reality. In *Proceedings of the 28th Annual ACM Symposium on User Interface Software & Technology* (UIST '15), ACM, New York, NY, USA, 405–415. DOI:https://doi.org/10.1145/2807442.2807497

[46] William Odom. 2010. "Mate, We Don't Need a Chip to Tell Us the Soil's Dry": Opportunities for Designing Interactive Systems to Support Urban Food Production. In *Proceedings of the 8th ACM Conference on Designing Interactive Systems* (DIS '10), ACM, New York, NY, USA, 232–235. DOI:https://doi.org/10.1145/1858171.1858211

[47] Jason Orlosky, Misha Sra, Kenan Bektaş, Huaishu Peng, Jeeeun Kim, Nataliya Kos'myna, Tobias Höllerer, Anthony Steed, Kiyoshi Kiyokawa, and Kaan Akşit. 2021. Telelife: The Future of Remote Living. *Frontiers in Virtual Reality* 2, (2021). Retrieved August 18, 2022 from https://www.frontiersin.org/articles/10.3389/frvir.2021.763340

[48] Sergio Orts-Escolano, Christoph Rhemann, Sean Fanello, Wayne Chang, Adarsh Kowdle, Yury Degtyarev, David Kim, Philip L. Davidson, Sameh Khamis, Mingsong Dou, Vladimir Tankovich, Charles Loop, Qin Cai, Philip A. Chou, Sarah Mennicken, Julien Valentin, Vivek Pradeep, Shenlong Wang, Sing Bing Kang, Pushmeet Kohli, Yuliya Lutchyn, Cem Keskin, and Shahram Izadi. 2016. Holoportation: Virtual 3D Teleportation in Real-time. In *Proceedings of the 29th Annual Symposium on User Interface Software and Technology* (UIST '16), ACM, New York, NY, USA, 741–754. DOI:https://doi.org/10.1145/2984511.2984517

[49] Sanela Osmanovic and Loretta Pecchioni. 2016. Beyond Entertainment: Motivations and Outcomes of Video Game Playing by Older Adults and Their Younger Family Members. *Games and Culture* 11, 1–2 (January 2016), 130–149. DOI:https://doi.org/10.1177/1555412015602819

[50] Aswati Panicker, Kavya Basu, and Chia-Fang Chung. 2020. Changing Roles and Contexts: Symbolic Interactionism in the Sharing of Food and Eating Practices between Remote, Intergenerational Family Members. *Proc. ACM Hum.-Comput. Interact.* 4, CSCW1 (May 2020), 43:1-43:19. DOI:https://doi.org/10.1145/3392848

[51] Tomislav Pejsa, Julian Kantor, Hrvoje Benko, Eyal Ofek, and Andrew Wilson. 2016. Room2Room: Enabling Life-Size Telepresence in a Projected Augmented Reality Environment. In *Proceedings of the 19th ACM Conference on Computer-Supported Cooperative Work & Social Computing* (CSCW '16), Association for Computing Machinery, New York, NY, USA, 1716–1725. DOI:https://doi.org/10.1145/2818048.2819965

[52] Melissa Escamilla Perez, Stephanie T. Jones, Sarah Priscilla Lee, and Marcelo Worsley. 2020. Intergenerational Making with Young Children. In *Proceedings of the FabLearn 2020 - 9th Annual Conference on Maker Education* (FabLearn '20), Association for Computing Machinery, New York, NY, USA, 68–73. DOI:https://doi.org/10.1145/3386201.3386225

[53] Daniela Petrelli and Steve Whittaker. 2010. Family memories in the home: contrasting physical and digital mementos. *Personal Ubiquitous Comput.* 14, 2 (February 2010), 153–169. DOI:https://doi.org/10.1007/s00779-009-0279-7

[54] Daniela Petrelli, Steve Whittaker, and Jens Brockmeier. 2008. AutoTopography: what can physical mementos tell us about digital memories? In *Proceedings of the SIGCHI Conference on Human Factors in Computing Systems* (CHI '08), Association for Computing Machinery, New York, NY, USA, 53–62. DOI:https://doi.org/10.1145/1357054.1357065

[55] Thammathip Piumsomboon, Gun A. Lee, Andrew Irlitti, Barrett Ens, Bruce H. Thomas, and Mark Billinghurst. 2019. On the Shoulder of the Giant: A Multi-Scale Mixed Reality Collaboration with 360 Video Sharing and Tangible Interaction. In *Proceedings of the 2019 CHI Conference on Human Factors in Computing Systems* (CHI '19), ACM, New York, NY, USA, 228:1-228:17. DOI:https://doi.org/10.1145/3290605.3300458

[56] Christina Pollalis, Elizabeth Joanna Minor, Lauren Westendorf, Whitney Fahnbulleh, Isabella Virgilio, Andrew L. Kun, and Orit Shaer. 2018. Evaluating Learning with Tangible and Virtual Representations of Archaeological Artifacts. In *Proceedings of the Twelfth International Conference on Tangible, Embedded, and Embodied Interaction* (TEI '18), Association for Computing Machinery, New York, NY, USA, 626–637. DOI:https://doi.org/10.1145/3173225.3173260

[57] Proximie. *How technology is boosting peer-to-peer collaboration in surgery*. Retrieved from https://www.proximie.com/how-technology-is-boosting-peer-to-peer-collaboration-in-surgery/

[58] Iulian Radu, Tugce Joy, and Bertrand Schneider. 2021. Virtual Makerspaces: Merging AR/VR/MR to Enable Remote Collaborations in Physical Maker Activities. In *Extended Abstracts of the 2021 CHI Conference on Human Factors in Computing Systems*. Association for Computing Machinery, New York, NY, USA, 1–5. Retrieved October 31, 2021 from https://doi.org/10.1145/3411763.3451561

[59] Hayes Raffle, Glenda Revelle, Koichi Mori, Rafael Ballagas, Kyle Buza, Hiroshi Horii, Joseph Kaye, Kristin Cook, Natalie Freed, Janet Go, and Mirjana Spasojevic. 2011. Hello, is grandma there? let's read! StoryVisit: family video chat and connected e-books. In *Proceedings of the SIGCHI Conference on Human Factors in Computing Systems* (CHI '11), Association for Computing Machinery, New York, NY, USA, 1195–1204. DOI:https://doi.org/10.1145/1978942.1979121

[60] Shwetha Rajaram, Franziska Roesner, and Michael Nebeling. 2021. Designing Privacy-Informed Sharing Techniques for Multi-User AR Experiences. *VR4Sec: 1st International Workshop on Security for XR and XR for Security* (January 2021). Retrieved August 18, 2022 from https://par.nsf.gov/biblio/10312789-designing-privacy-informed-sharing-techniques-multi-user-ar-experiences

[61] Logan Reis, Kathryn Mercer, and Jennifer Boger. Technologies for fostering intergenerational connectivity and relationships: Scoping review and emergent concepts | Elsevier Enhanced Reader. DOI:https://doi.org/10.1016/j.techsoc.2020.101494

[62] Selma Rizvic, Vensada Okanovic, Irfan Prazina, and Aida Sadzak. 2016. *4D Virtual Reconstruction of White Bastion Fortress*. The Eurographics Association. DOI:https://doi.org/10.2312/gch20161387

[63] Selma Rizvic and Irfan Prazina. 2015. Taslihan Virtual Reconstruction - Interactive Digital Story or a Serious Game. In *2015 7th International Conference on Games and Virtual Worlds for Serious Applications (VS-Games)*, 1–2. DOI:https://doi.org/10.1109/VS-GAMES.2015.7295786

[64] Jim Scheibmeir. 2021. Quality Attributes of Digital Twins. PhD Thesis. Colorado State University.

[65] Claudia Schlegel, Alain Geering, and Uwe Weber. 2021. Learning in virtual space: an intergenerational pilot project. *GMS J Med Educ* 38, 2 (February 2021), Doc37. DOI:https://doi.org/10.3205/zma001433

[66] Minyeong Seo, Hansu Lee, Seungmi Choi, Suhyun Jo, Heejae Jung, Subin Park, and Hyunggu Jung. 2019. Exploring Experiences of Virtual Reality among Young and Older Adults in a Subway Fire Scenario: a Pilot Study. In *25th ACM Symposium on Virtual Reality Software and Technology* (VRST '19), Association for Computing Machinery, New York, NY, USA, 1–2. DOI:https://doi.org/10.1145/3359996.3364788

[67] Elisa Serafinelli and Lauren Alex O'Hagan. 2022. Drone views: a multimodal ethnographic perspective. *Visual Communication* (May 2022), 14703572211065092. DOI:https://doi.org/10.1177/14703572211065093

[68] Mike Shafto, M Conroy, R Doyle, E Glaessgen, C Kemp, J LeMoigne, and L Wang. 2010. *Modeling, Simulation, Information Technology and Processing Roadmap*.

[69] Rita Shewbridge, Amy Hurst, and Shaun K. Kane. 2014. Everyday making: identifying future uses for 3D printing in the home. In *Proceedings of the 2014 conference on Designing interactive systems* (DIS '14), Association for Computing Machinery, New York, NY, USA, 815–824. DOI:https://doi.org/10.1145/2598510.2598544

[70] Adalberto L. Simeone and Eduardo Velloso. 2015. Substitutional reality: bringing virtual reality home. *XRDS* 22, 1 (November 2015), 24–29. DOI:https://doi.org/10.1145/2810044

[71] Mike Sinclair, Eyal Ofek, Mar Gonzalez-Franco, and Christian Holz. 2019. CapstanCrunch: A Haptic VR Controller with User-supplied Force Feedback. In *Proceedings of the 32nd Annual ACM Symposium on User Interface Software and Technology* (UIST '19), Association for Computing Machinery, New York, NY, USA, 815–829. DOI:https://doi.org/10.1145/3332165.3347891

[72] Harrison Jesse Smith and Michael Neff. 2018. Communication Behavior in Embodied Virtual Reality. In *Proceedings of the 2018 CHI Conference on Human Factors in Computing Systems* (CHI '18), ACM, New York, NY, USA, 289:1-289:12. DOI:https://doi.org/10.1145/3173574.3173863

[73] Misha Sra and Chris Schmandt. 2016. Bringing real objects, spaces, actions, and interactions into social VR. In *2016 IEEE Third VR International Workshop on Collaborative Virtual Environments (3DCVE)*, 16–17. DOI:https://doi.org/10.1109/3DCVE.2016.7563561

[74] Scott Stein. Google 3D animals: How to conjure AR animals with Google search and more. *CNET*. Retrieved September 2, 2022 from https://www.cnet.com/tech/services-and-software/google-3d-animals-how-to-use-cool-ar-feature-at-home-list-of-objects/

[75] Paul Strohmeier and Kasper Hornbæk. 2017. Generating Haptic Textures with a Vibrotactile Actuator. In *Proceedings of the 2017 CHI Conference on Human Factors in Computing Systems* (CHI '17), Association for Computing Machinery, New York, NY, USA, 4994–5005. DOI:https://doi.org/10.1145/3025453.3025812

[76] Hongling Sun, Yue Liu, Zhenliang Zhang, Xiaoxu Liu, and Yongtian Wang. 2018. Employing Different Viewpoints for Remote Guidance in a Collaborative Augmented Environment. In *Proceedings of the Sixth International Symposium of Chinese CHI* (ChineseCHI '18), Association for Computing Machinery, New York, NY, USA, 64–70. DOI:https://doi.org/10.1145/3202667.3202676

[77] Anthony Tang, Omid Fakourfar, Carman Neustaedter, and Scott Bateman. 2017. Collaboration with 360° Videochat: Challenges and Opportunities. In *Proceedings of the 2017 Conference on Designing Interactive Systems* (DIS '17), Association for Computing Machinery, New York, NY, USA, 1327–1339. DOI:https://doi.org/10.1145/3064663.3064707

[78] Andre Taulien, Anika Paulsen, Tim Streland, Benedikt Jessen, Stefan Wittke, and Michael Teistler. 2019. A Mixed Reality Environmental Simulation to Support Learning about Maritime Habitats: An Approach to Convey Educational Knowledge With a Novel User Experience. In *Proceedings of Mensch und Computer 2019* (MuC'19), Association for Computing Machinery, New York, NY, USA, 921–925. DOI:https://doi.org/10.1145/3340764.3345382

[79] Theophilus Teo, Louise Lawrence, Gun A. Lee, Mark Billinghurst, and Matt Adcock. 2019. Mixed Reality Remote Collaboration Combining 360 Video and 3D Reconstruction. In *Proceedings of the 2019 CHI Conference on Human Factors in Computing Systems* (CHI '19), Association for Computing Machinery, New York, NY, USA, 1–14. DOI:https://doi.org/10.1145/3290605.3300431

[80] Theophilus Teo, Mitchell Norman, Gun A. Lee, Mark Billinghurst, and Matt Adcock. 2020. Exploring interaction techniques for 360 panoramas inside a 3D reconstructed scene for mixed reality remote collaboration. *J Multimodal User Interfaces* 14, 4 (December 2020), 373–385. DOI:https://doi.org/10.1007/s12193-020-00343-x

[81] Arnout Terpstra, Alexander P. Schouten, Alwin de Rooij, and Ronald E. Leenes. 2019. Improving privacy choice through design: How designing for reflection could support privacy self-management. *First Monday* (June 2019). DOI:https://doi.org/10.5210/fm.v24i7.9358

[82] Balasaravanan Thoravi Kumaravel, Fraser Anderson, George Fitzmaurice, Bjoern Hartmann, and Tovi Grossman. 2019. Loki: Facilitating Remote Instruction of Physical Tasks Using Bi-Directional Mixed-Reality Telepresence. In *Proceedings of the 32nd Annual ACM Symposium on User Interface Software and Technology* (UIST '19), Association for Computing Machinery, New York, NY, USA, 161–174. DOI:https://doi.org/10.1145/3332165.3347872

[83] Rundong Tian, Sarah Sterman, Ethan Chiou, Jeremy Warner, and Eric Paulos. 2018. MatchSticks: Woodworking through Improvisational Digital Fabrication. In *Proceedings of the 2018 CHI Conference on Human Factors in Computing Systems* (CHI '18), Association for Computing Machinery, New York, NY, USA, 1–12. DOI:https://doi.org/10.1145/3173574.3173723

[84] Priscilla Ulguim. 2018. Models and Metadata: The Ethics of Sharing Bioarchaeological 3D Models Online. *Arch* 14, 2 (August 2018), 189–228. DOI:https://doi.org/10.1007/s11759-018-9346-x

[85] Cheng Yao Wang, Sandhya Sriram, and Andrea Stevenson Won. 2021. Shared Realities: Avatar Identification and Privacy Concerns in Reconstructed Experiences. *Proc. ACM Hum.-Comput. Interact.* 5, CSCW2 (October 2021), 337:1-337:25. DOI:https://doi.org/10.1145/3476078

[86] Christian Weichel, Manfred Lau, David Kim, Nicolas Villar, and Hans W. Gellersen. 2014. MixFab: a mixed-reality environment for personal fabrication. In *Proceedings of the SIGCHI Conference on Human Factors in Computing Systems* (CHI '14), Association for Computing Machinery, New York, NY, USA, 3855–3864. DOI:https://doi.org/10.1145/2556288.2557090

[87] Maximiliane Windl, Niels Henze, Albrecht Schmidt, and Sebastian S. Feger. 2022. Automating Contextual Privacy Policies: Design and Evaluation of a Production Tool for Digital Consumer Privacy Awareness. In *Proceedings of the 2022 CHI Conference on Human Factors in Computing Systems* (CHI '22), Association for Computing Machinery, New York, NY, USA, 1–18. DOI:https://doi.org/10.1145/3491102.3517688

[88] Yufei Ye, Abhinav Gupta, and Shubham Tulsiani. 2022. What's in your hands? 3D Reconstruction of Generic Objects in Hands. (April 2022). DOI:https://doi.org/10.48550/arXiv.2204.07153

[89] Xenophon Zabulis, Carlo Meghini, Nikolaos Partarakis, Cynthia Beisswenger, Arnaud Dubois, Maria Fasoula, Vito Nitti, Stavroula Ntoa, Ilia Adami, Antonios Chatziantoniou, Valentina Bartalesi, Daniele Metilli, Nikolaos Stivaktakis, Nikolaos Patsiouras, Paraskevi Doulgeraki, Effie Karuzaki, Evropi Stefanidi, Ammar Qammaz, Danae Kaplanidi, and George Galanakis. 2020. Representation and Preservation of Heritage Crafts. *Sustainability* 12, (February 2020), 1461. DOI:https://doi.org/10.3390/su12041461

[90] Yaying Zhang, Brennan Jones, Sean Rintel, and Carman Neustaedter. 2021. XRmas: Extended Reality Multi-Agency Spaces for a Magical Remote Christmas. Retrieved from https://www.microsoft.com/en-us/research/publication/xrmas-extended-reality-multi-agency-spaces-for-a-magical-remote-christmas/

[91] Z. Zhou, A.D. Cheok, S.P. Lee, L.N. Thang, C.K. Kok, W.Z. Ng, Y.K. Cher, M.L. Pung, and Y. Li. 2005. Age Invader: human media for natural social-physical inter-generational interaction with elderly and young. In *Proceedings of the 2005 International Conference on Active Media Technology, 2005. (AMT 2005).*, 203–204. DOI:https://doi.org/10.1109/AMT.2005.1505308

[92] Jakob Zillner, Erick Mendez, and Daniel Wagner. 2018. Augmented Reality Remote Collaboration with Dense Reconstruction. In *2018 IEEE International Symposium on Mixed and Augmented Reality Adjunct (ISMAR-Adjunct)*, 38–39. DOI:https://doi.org/10.1109/ISMAR-Adjunct.2018.00028

[93] 2014. *NGA Garden to Table Report*. National Gardening Association. Retrieved from https://garden.org/special/pdf/2014-NGA-Garden-to-Table.pdf

[94] 2015. *Introducing YardMap: Citizen-Scientists Make a Difference*. Retrieved December 12, 2022 from https://www.youtube.com/watch?v=-MFR20oN_5w

[95] 2018. *Osso VR, Virtual Reality Surgical Training Platform*. Retrieved from https://ossovr.com/

[96] 2018. *Facebook 3D VR Reconstruction for Photo Memories*. Retrieved September 6, 2022 from https://www.youtube.com/watch?v=Ur1Z72_LyTM

[97] 2019. *XR Access Symposium Report*. Retrieved from https://docs.google.com/document/d/131eLNGES3_2M5_roJacWlLhX-nHZqghNhwUgBF5lJaE/edit

[98] 2019. *Facebook Reality Labs: LiveMaps | Oculus Connect 6*. Retrieved August 30, 2022 from https://www.youtube.com/watch?v=JTa8zn0RNVM

[99] 2019. Voice guidance in Maps, built for people with impaired vision. *Google*. Retrieved August 30, 2022 from https://blog.google/products/maps/better-maps-for-people-with-vision-impairments/

[100] 2020. *Remote Assist, Microsoft Dynamics 365*. Retrieved from https://dynamics.microsoft.com/en-us/mixed-reality/remote-assist/

[101] 2021. An Era of Digital Humans: Pushing the Envelope of Photorealistic Digital Character Creation. *NVIDIA Technical Blog*. Retrieved September 11, 2022 from https://developer.nvidia.com/blog/an-era-of-digital-humans-pushing-the-envelope-of-photorealistic-digital-character-creation/

[102] Building Mixed Reality with Presence Platform: A New Dimension for Work and Fun. Retrieved September 11, 2022 from https://www.oculus.com/blog/building-mixed-reality-with-presence-platform-a-new-dimension-for-work-and-fun/

[103] Pushing state-of-the-art in 3D content understanding. Retrieved September 11, 2022 from https://ai.facebook.com/blog/pushing-state-of-the-art-in-3d-content-understanding/
[104] Agisoft Metashape: Agisoft Metashape. Retrieved November 29, 2022 from https://www.agisoft.com/
[105] networked-aframe/networked-aframe: A web framework for building multi-user virtual reality experiences. Retrieved September 10, 2022 from https://github.com/networked-aframe/networked-aframe
[106] Introducing the Niantic Planet-Scale AR Alliance: Bringing the Mobile Industry Together Towards the 5G Future of Consumer AR. Retrieved August 30, 2022 from https://nianticlabs.com/news/niantic-planet-scale-ar-alliance-5g/
[107] Ethical Considerations in Augmented Reality Applications - ProQuest. Retrieved September 7, 2022 from https://www.proquest.com/openview/563a017acbaddcfef2b659a9b6d3457f/1?cbl=1976356&pq-origsite=gscholar
[108] Memento. *Memento*. Retrieved September 6, 2022 from https://www.mementovr.com